\documentclass[twocolumn, showpacs,preprintnumbers,amsmath,amssymb,nofootinbib]{revtex4}
\usepackage{graphicx}% Include figure files
\usepackage{dcolumn}% Align table columns on decimal point
\usepackage{bm}% bold math
\usepackage{epsfig}
\usepackage{color}

\usepackage{color}
\usepackage{hyperref}

\hypersetup{
    colorlinks=true,
    linkcolor=red,
    citecolor=blue,
}

\newcommand{\wl}{{\rm WL}}
\newcommand{\vp}{{\rm v}}

\def\fun#1#2{\lower3.6pt\vbox{\baselineskip0pt\lineskip.9pt
  \ialign{$\mathsurround=0pt#1\hfil##\hfil$\crcr#2\crcr\sim\crcr}}}
\def\simgt{\mathrel{\lower0.6ex\hbox{$\buildrel {\textstyle >}
 \over {\scriptstyle \sim}$}}}
\def\simlt{\mathrel{\lower0.6ex\hbox{$\buildrel {\textstyle <}
 \over {\scriptstyle \sim}$}}}

\input epsf

\newcommand{\hompc}{\,h\,{\rm Mpc}^{-1}}
\newcommand{\mpcoh}{\,h^{-1}\,{\rm Mpc}}

\def\be{\begin{equation}}
\def\ee{\end{equation}}
\def\ba{\begin{eqnarray}}
\def\ea{\end{eqnarray}}

\def\nn{\nonumber}
\begin{document}

\preprint{}

\title{Complementarity of Weak Lensing and Peculiar Velocity Measurements \\ in Testing General Relativity}

\author{Yong-Seon Song$^{1,2}$\email{ysong@kias.re.kr}, Gong-Bo Zhao$^2$, David Bacon$^2$, Kazuya Koyama$^2$, Robert C Nichol$^2$, Levon Pogosian$^3$}
\affiliation{$^1$Korea Institute for Advanced Study, Dongdaemun-gu, Seoul 130-722, Korea \\
$^2$Institute of Cosmology \& Gravitation, University of Portsmouth, Dennis Sciama Building, Portsmouth, PO1 3FX, United Kingdom \\
$^3$Department of Physics, Simon Fraser University, Burnaby, BC, V5A 1S6, Canada}

%\date{\today}

\begin{abstract}

We explore the complementarity of weak lensing and galaxy peculiar velocity measurements to better constrain modifications to General Relativity. We find no evidence for deviations from GR on cosmological scales from a combination of peculiar velocity measurements (for Luminous Red Galaxies in the Sloan Digital Sky Survey) with weak lensing measurements (from the CFHT Legacy Survey). We provide a Fisher error forecast for a Euclid-like space-based survey including both lensing and peculiar velocity measurements, and show that the expected constraints on modified gravity will be at least an order of magnitude better than with present data, i.e. we will obtain $\simeq5\%$ errors on the modified gravity parametrization described here. We also present a model--independent method for constraining modified gravity parameters using tomographic peculiar velocity information, and apply this methodology to the present dataset.
\end{abstract}

\pacs{draft}


\maketitle

\section{Introduction}

The discovery of the cosmic acceleration \cite{Perlmutter:1998np, Riess:1998cb} a decade ago has forced cosmologists to modify their simple picture of the Universe, i.e. a universe dominated only by matter and described solely by General Relativity (GR). Over the last ten years, physicists have suggested two main avenues for explaining the late--time acceleration of the Universe. The first is the introduction of Dark Energy (DE), with an effective negative pressure, which dominates the late--time energy density of the Universe thus causing the acceleration. An alternative explanation is to modify the law of gravity on large scales thus altering the predicted expansion history of the Universe to be in line with the observations. In this paper, we focus on this latter explanation for the cosmic acceleration and present combinations of observables that can be used to test the validity of GR on cosmological scales.

General Relativity is a metric theory of gravity that can describe the relationship between matter perturbations, gravitational potential and space curvature perturbations. Several authors have shown that by combining various probes of the large--scale structure in the Universe, it is possible to test the relationship between these quantities which, in the linear regime, can generally be described by two functions of time and scale \cite{Jain:2007yk, Song:2008vm, Song:2008xd, Zhao:2008bn, Zhao:2009fn, Pogosian:2010tj,Kunz:2006ca,Amendola:2007rr} (see \cite{Jain:2010ka} for a review and references therein). For example, several recent attempts have been made to constrain parameters associated with these two functions using the latest weak lensing measurements \cite{Daniel:2010ky, Bean:2010zq, Zhao:2010dz, Daniel:2010yt} and these current analyses appear to be consistent with GR once systematic errors are taken into account. However, these current constraints on Modified Gravity (MG) parameters are still weak, especially as there are degeneracies between the parameters that cannot be broken by weak lensing measurements alone.

We show in this paper that these degeneracies can be broken through a combination of weak lensing (WL) and peculiar velocity (PV) measurements (see \cite{Guzik:2009cm} for an earlier approach). This is motivated by the fact that the WL experiments probe the lensing potential, which determines the trajectories of photons through the Universe, while PV measurements probe the gravitational potential that governs the dynamics of galaxies independent of their galaxy bias. A combination of these two observables can, in principle, simultaneously measure the MG parameters used to describe the possible modifications to the relationships between the metric perturbations defined in GR \cite{Jain:2007yk, Song:2008vm}.

This paper is organized as follows. In Section \ref{formalism}, we introduce the two MG parameters that describe the relationships between the GR metric perturbations. In Section \ref{current}, we use current observational data to present constraints on these MG parameters. In Section \ref{forecast}, we perform a Fisher matrix error forecast for a future Euclid-like space--based DE mission, while in section \ref{independent} we discuss a new method for extracting model--independent information from PV measurements. We conclude in Section \ref{conclusion}.

\section{Modified Gravity Parameters}
\label{formalism}
Linear metric perturbations around a background Friedmann universe are described by the line element in the Newtonian gauge
\be\label{eq:metric}
ds^2=-(1+2\Psi)dt^2+(1+2\Phi)a^2\delta_{ij}dx^idx^j\,,
\ee
where $\Phi$ and $\Psi$ denote the space curvature perturbation and the gravitational potential respectively. The dynamics of pressureless matter fluctuations are determined by energy-momentum conservation, and are given by
\ba\label{eq:conti}
\dot{\delta}+\frac{\theta}{a} &=& 0, \\
\dot{\theta}+H\theta&=&\frac{k^2}{a}\Psi\,,
\label{eq:Euler}
\ea
where $k$ is the wavenumber of the perturbations, $\delta$ is the energy density contrast, $\theta$ is the divergence of the velocity perturbations, and the overdot denotes the derivative with respect to the physical time $t$. In this paper, we
only consider perturbations on sub-horizon, but linear, scales (i.e. $H\ll k/a\lesssim0.1 $ h/Mpc).

Two additional equations, which need to be provided by a theory of gravity, are required to solve the evolution equations for the four scalar perturbations given above ($\Phi, \Psi, \delta$ and $\theta$). In this paper, we use the following modified gravitational equations to parametrize possible modifications from the expected GR relations between these perturbations (see ~\cite{Pogosian:2010tj}, and references therein, for a discussion of the various possible parametrizations),
\ba
\label{eq:MG_para1}k^2\Psi &=& -4\pi G a^2\mu(k, a) \rho \delta, \\
\label{eq:MG_para2}k^2 (\Phi - \Psi) &=& 8 \pi G a^2 \Sigma(k, a) \rho \delta,
\ea
where $G$ is the Newton constant measured in a Cavendish-type experiment. Here, the function
$\mu$ characterizes a modification of Newton's constant in both space and time, and the $\Sigma$ function describes a modification to the lensing potential $\Phi - \Psi$, again with space and time. Since $\mu$ affects the gravitational potential $\Psi$, it changes the growth rate of density fluctuations $\delta$ via Eqns.~(\ref{eq:conti}) and (\ref{eq:Euler}). Weak lensing measurements probe the lensing potential, and are thus affected by both $\Sigma$ and $\mu$ (via $\delta$). Therefore there is a degeneracy in the constraints on $\Sigma$ and $\mu$ from WL.

On the other hand, peculiar velocities of galaxies are determined by the gravitational potential $\Psi$ through the Euler equation, Eqn.~(\ref{eq:Euler}), and are thus affected by $\mu$ but not by $\Sigma$. Therefore, the combination of WL and PV measurements can break the WL degeneracy between $\Sigma$ and $\mu$, and allow cosmologists to probe these two functions separately. Furthermore, ~\cite{Song:2010rm} recently studied the theoretical priors on $\mu$ and $\Sigma$ from scalar-tensor gravity theories, clustering dark energy models and interacting dark energy models, and showed that each of these theories has a distinct path in the $\mu$--$\Sigma$ parameter space thus providing the opportunity to distinguish between these possible theoretical models.

For the purpose of highlighting the key features of the combined constraints, in the following sections we use a simplified parameterization of $\Sigma$ and $\mu$ given by
\begin{equation}
\Sigma = 1 + \Sigma_s a^s, \quad
\mu  = 1 + \mu_s a^s,
\label{parametrization}
\end{equation}
where $s$ specifies the power of the assumed time variation of $\Sigma$ and $\mu$, and $\Sigma_s$ and $\mu_s$ are constants.
%The subscript $s$ makes it clear that the constant was used in conjunction with a corresponding time-dependence.
In Section \ref{forecast} we will consider two cases, $s=1$ (the linear model) and $s=3$ (the cubic model). The linear model is motivated by DGP \cite{Dvali:2000hr, Koyama:2005kd, Song:2006sa},
while the cubic model is motivated by general plausibility arguments that $\mu$ may change
in proportion to matter density \cite{Linder:2007hg, Daniel:2010ky}.

The parametrization in Eqn.~(\ref{parametrization}) assumes that $\Sigma$ and $\mu$ are scale independent. It should be emphasized that this is simply a restriction that arises when using the current data sets. It has been shown that the WL data is far more sensitive to scale--dependent modifications to GR \cite{Zhao:2009fn, Pogosian:2010tj, Zhao:2010dz}. Moreover, the PV observations we use in Section \ref{current} {\it assume} that the growth rate is scale--independent \cite{Song:2010kq}, and therefore this data cannot be used for tests of scale--dependent models. In order to make a consistent comparison between current errors and future forecasts, we also do not consider the scale--dependent models in our future forecasts in Section \ref{forecast}. A comprehensive forecast of general scale and time dependent MG constraints from future PV measurements will be presented separately \cite{future}.

\begin{figure*}[t]
\includegraphics[width=7cm]{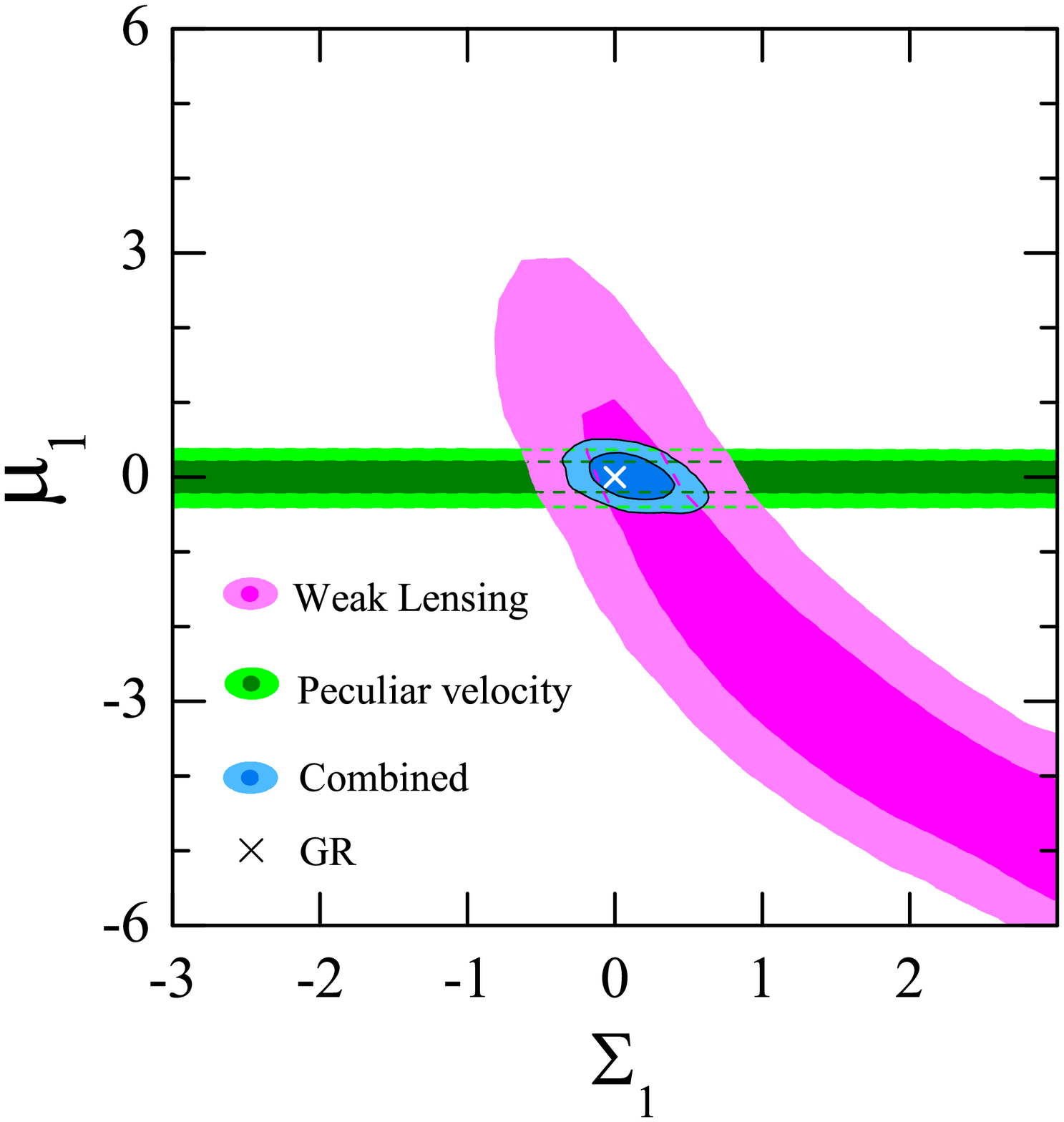}
\includegraphics[width=7cm]{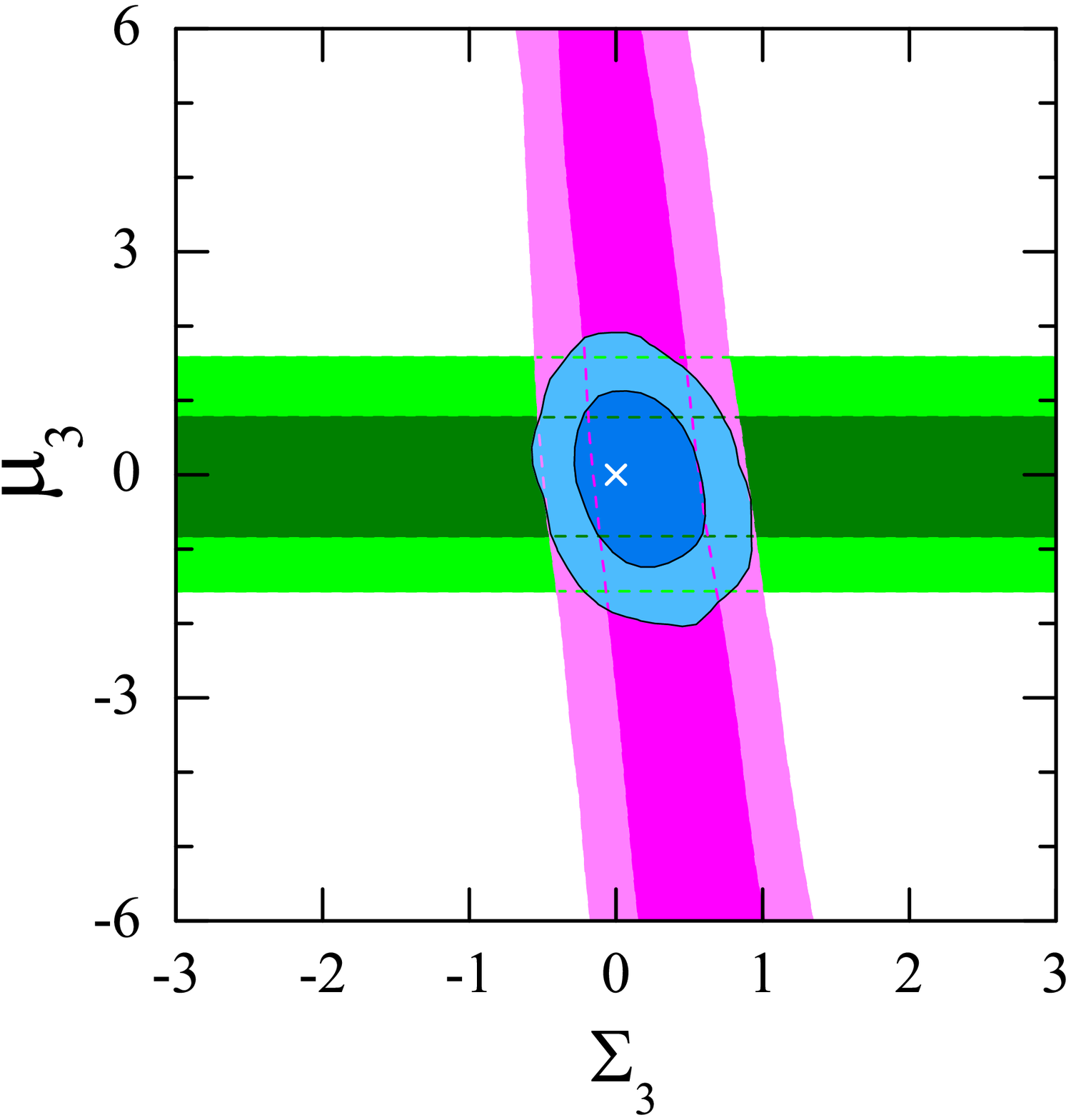}
\caption{The constraints on $\Sigma_s$ and $\mu_s$ from the latest observational data; $s=1$ linear model on the left and the $s=3$ cubic model on the right. The inner dark-shaded contours are the 68\% confidence region, while the outer lighter-shaded contours are the 95\% confidence region. The horizontal green bands show the peculiar velocity constraints, while the magenta bands show the weak lensing constraints. The central blue contours show the constraints possible when the two datasets are analysed together. The cross symbol shows the expected parameterization for General Relativity.}\label{fig:current}
\end{figure*}

\section{Current Constraints}
\label{current}

In this section, we present the current constraints on $\mu$ and $\Sigma$ defined in Section \ref{formalism} from the weak lensing measurements of the CFHT Legacy Survey (CFHTLS) \cite{Fu:2007qq, Kilbinger:2008gk}, and PV measurements obtained from Sloan Digital Sky Survey (SDSS) Luminous Red Galaxies (LRGs) selected from their Data Release Seven (DR7) \cite{Song:2010kq}.

\subsection{Weak lensing measurements}
The WL shear power spectrum is equal to the convergence power spectrum given by
\cite{Zhao:2008bn}
\be
\label{eq:Cl_shear} C_{\ell}^{\kappa} =\frac{36\pi}{25}\int{\rm dln}
k\Delta_{\mathcal{R}}^2\left[\int{\rm d}zW(z)j_{\ell}[k
\chi(z)](\Phi-\Psi)(k,z)\right]^2,
\ee
where $\Delta_{\mathcal{R}}^2$ denotes the primordial curvature power
spectrum, $\chi(z)$ is the comoving distance and
the window function for WL is defined as $ W(z)
=\int_z^{\infty}{\rm d}z'n(z')[\chi(z')/\chi(z)-1]$.
Given $C_{\ell}^{\kappa}$, we can obtain the E-mode component of the
shear correlations $\xi_E$ via \cite{Crittenden:2000au}, 
\be
 \xi_E(\theta)=\frac{1}{2\pi}\int_0^{\infty}{\rm d}\ell
{\ell}C_{\ell}^{\kappa}{ J_0}(\ell\theta),
\ee
where $J_0$ is the Bessel function of the first kind.

For the WL observable, we use the $\xi_E$ data from the CFHTLS-Wide survey, presented in \cite{Fu:2007qq}; this is derived from shear measurements for $2\times10^6$ galaxies with magnitudes $21.5<i_{AB}<24.5$ in a 35 sq. deg. effective area. We note that there is a modest systematic effect in the CFHTLS data, due to residual field-to-field variations in shear estimation on the scale of
the camera field of view \cite{Zhao:2010dz}.

We estimate the data covariance matrix using the Horizon
simulation as described in \cite{Beynon:2010}, section 3; we measure the covariance of correlation functions in 75 patches of area 16 deg$^2$, and estimate the combined covariance including shape noise and large-scale structure covariance using the results of \cite{hartlap}. To model the galaxy distribution function $n(z)$ for the CFHTLS sample,
we follow \cite{Fu:2007qq} in using the parametric form 
\be\label{eq:nz} n(z)\propto\frac{z^A+z^{AB}}{z^B+C},\ee where $A,B$ and $C$ are nuisance parameters to be marginalised over.  We wish to avoid strongly non-linear scales in modified gravity, as these cannot be treated
properly without using N-body simulations; we therefore remove from our analysis the $\xi_E$ data in the highly non-linear regime, namely $\theta<30$ arcmins.

\subsection{Peculiar velocity measurements}
The coherent peculiar motions of galaxies can be statistically estimated through the measurement, and modeling, of the large--scale redshift-space distortions \cite{Hamilton:1997zq}.
The radial peculiar velocities can be expressed as a one--dimensional velocity dispersion, $\sigma_v$, as discussed and defined in \cite{Song:2010bk, Song:2010kq}, for which
\be\label{eq:sigmav}
\sigma^2_v=\frac{1}{6\pi^2}\int P_{\Theta\Theta}(k,a)dk\,,
\ee
where $P_{\Theta\Theta}$ is the three--dimensional power spectrum of $\Theta=\theta/aH$, and $\sigma_v$ has units of $\mpcoh$. In \cite{Song:2010kq} the one--dimensional velocity dispersion of galaxies was determined from the two--dimensional two--point correlation function of SDSS DR7 LRGs, and was found to have the values  $\sigma_v=3.01^{+0.45}_{-0.46} \mpcoh$  at a mean redshift of $z=0.25$, and $\sigma_v=3.69^{+0.47}_{-0.47} \mpcoh$  at a mean redshift of $z=0.38$. We will therefore use these values of $sigma_v$ as our PV observable. We should note that these measurements assume a flat $\Lambda$CDM background, i.e. a dark energy equation of state fixed at $w=-1$. This is a valid constraint in our study, since the background expansion is known to be close to $\Lambda$CDM, and generically a modified gravity can mimic this expansion while having a different structure growth history. Therefore the constraints on modified gravity in this study come from  structure growth.

\begin{table*}[t]
\begin{tabular}{c|c|c|c|c|}
   \cline{2-5}
   % after \\: \hline or \cline{col1-col2} \cline{col3-col4} ...
    & \multicolumn{2}{c|}{s=1} &
\multicolumn{2}{c|}{s=3} \\  \cline{2-5}
    & $\mu_1$ & $\Sigma_1$ & $\mu_3$ & $ \Sigma_3$ \\  \hline
    \multicolumn{1}{|c|} {WL} &
$[-6.3,1.9]$ & $>-0.4$ & Unconstrained & $0.32\pm0.48^{+0.91}_{-0.84}$ \\  \hline
    \multicolumn{1}{|c|} {PV} &
$0.06\pm0.20\pm0.40$ & Unconstrained & $-0.03\pm0.80^{+1.6}_{-1.5}$ & Unconstrained \\  \hline
    \multicolumn{1}{|c|} {PV+WL} &
$-0.002\pm0.20\pm0.39$ & $0.11\pm0.19^{+0.40}_{-0.36}$ & $-0.08^{+0.77}
_{-0.78}\pm1.5$ & $0.17^{+0.29+0.59}_{-0.28-0.54}$ \\  \hline
\end{tabular}
\caption{We present the mean, 68\%, and 95\% confidence limits for the modified gravity parameters shown in Eqn. 6 for both $s=1$ and $s=3$. We provide constraints based on using the weak lensing (WL) and peculiar
velocity (PV) data alone, as well as the combination of these two measurements (PV+WL). We only provide 95\% confidence limits for the WL--only constraints on $\mu_1$ and $\Sigma_1$.}
\label{table1}
\end{table*}

\subsection{Current constraints on $\mu_s$ and $\Sigma_s$}
In this section, we present the current constraints on the MG parameters discussed in Section \ref{formalism} from the available WL and PV data discussed above. In detail, we vary the following set of parameters,
\be
\label{eq:paratriz} {\bf P} \equiv (\omega_{b}, \omega_{c},
\Theta_{s}, \tau, n_s, A_s, \mathcal{N}, \mu_s, \Sigma_s),
\ee
where
$\omega_{b}\equiv\Omega_{b}h^{2}$ and
$\omega_{c}\equiv\Omega_{c}h^{2}$ are the physical baryon and cold
dark matter densities relative to the critical density respectively,
$\Theta_{s}$ is the ratio (multiplied by 100) of the sound
horizon to the angular diameter distance at decoupling, $\tau$
denotes the optical depth to re-ionization, $n_s$ and $A_s$ are the
primordial power spectral index and amplitude respectively, and $\Sigma_s$ and $\mu_s$ are the MG parameters for our scale-independent parametrization from Eqn.~(\ref{parametrization}). We also vary, and marginalize over,
several nuisance parameters denoted by $\mathcal{N}$ when performing our likelihood analysis, including three parameters $(A, B, C)$ for WL data, associated with the galaxy distribution $n(z)$ in Eqn. (\ref{eq:nz}), and one for supernovae, which accounts for the calibration uncertainty in measuring the supernova intrinsic luminosity. Finally, we assume a flat $\Lambda$CDM background cosmology (as discussed in section B).

Given the set of cosmological parameters ${\bf P}$ in
Eqn.~(\ref{eq:paratriz}), we calculate the expected observables including
the CMB shift parameters \cite{Komatsu:2010fb}, the luminosity
distance for supernovae, the growth factor for PV ($\Theta$) and the E-mode component of the weak lensing shear ($\xi_E$) using {\tt MGCAMB} \cite{Zhao:2008bn}.
We then constrain the model parameters by comparing these predictions with
the PV and WL data discussed above, as well as available SNe from the UNION-2 sample \cite{Amanullah:2010vv} and CMB shift parameters derived from WMAP seven year data \cite{Komatsu:2010fb}, using a version of the Markov Chain Monte Carlo (MCMC) package {\tt CosmoMC}~\cite{Cosmomc, Lewis:2002ah} modified to include our extra parameters. 

\begin{figure*}[t]
\includegraphics[width= 7cm]{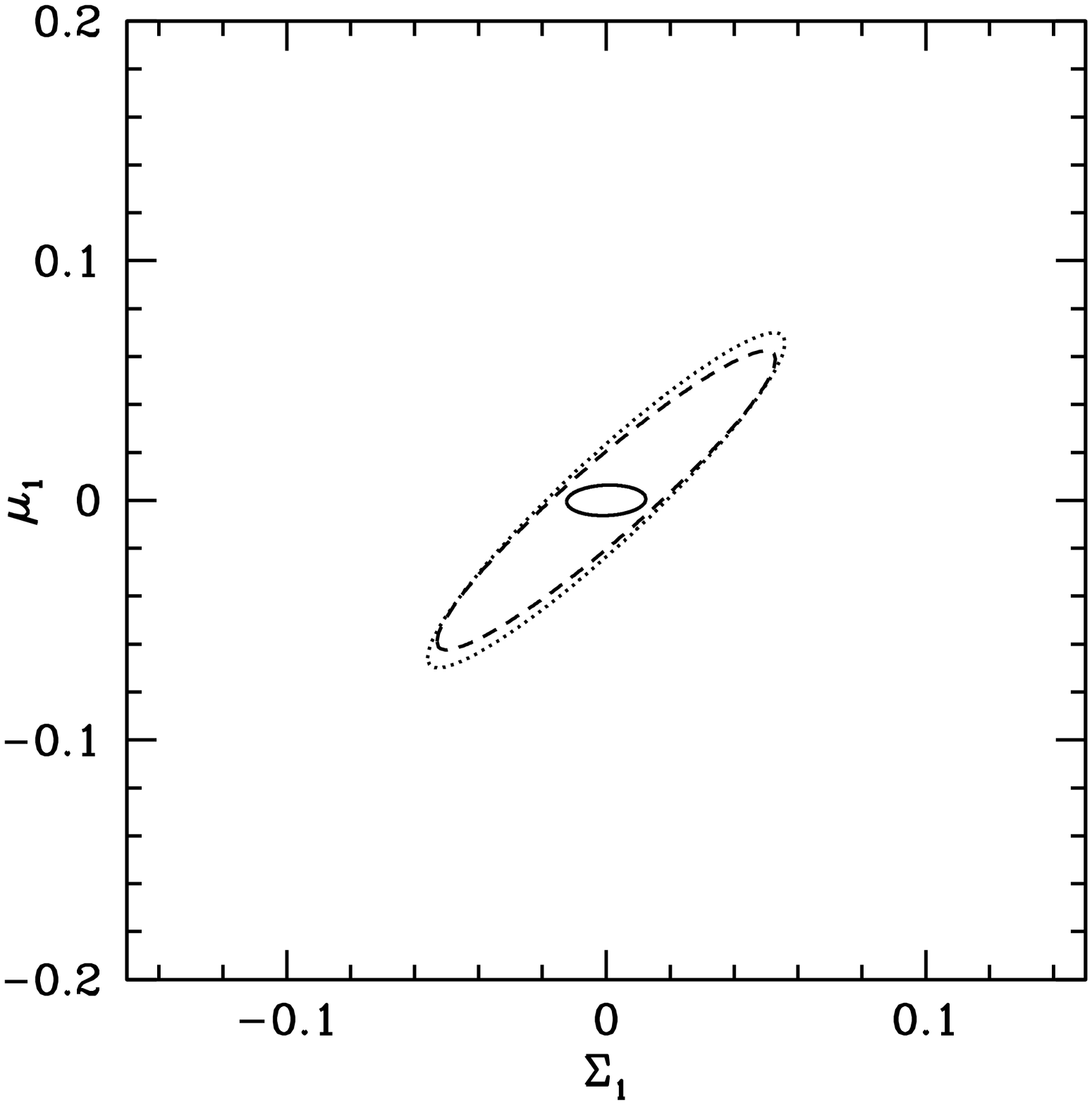}
\includegraphics[width= 7cm]{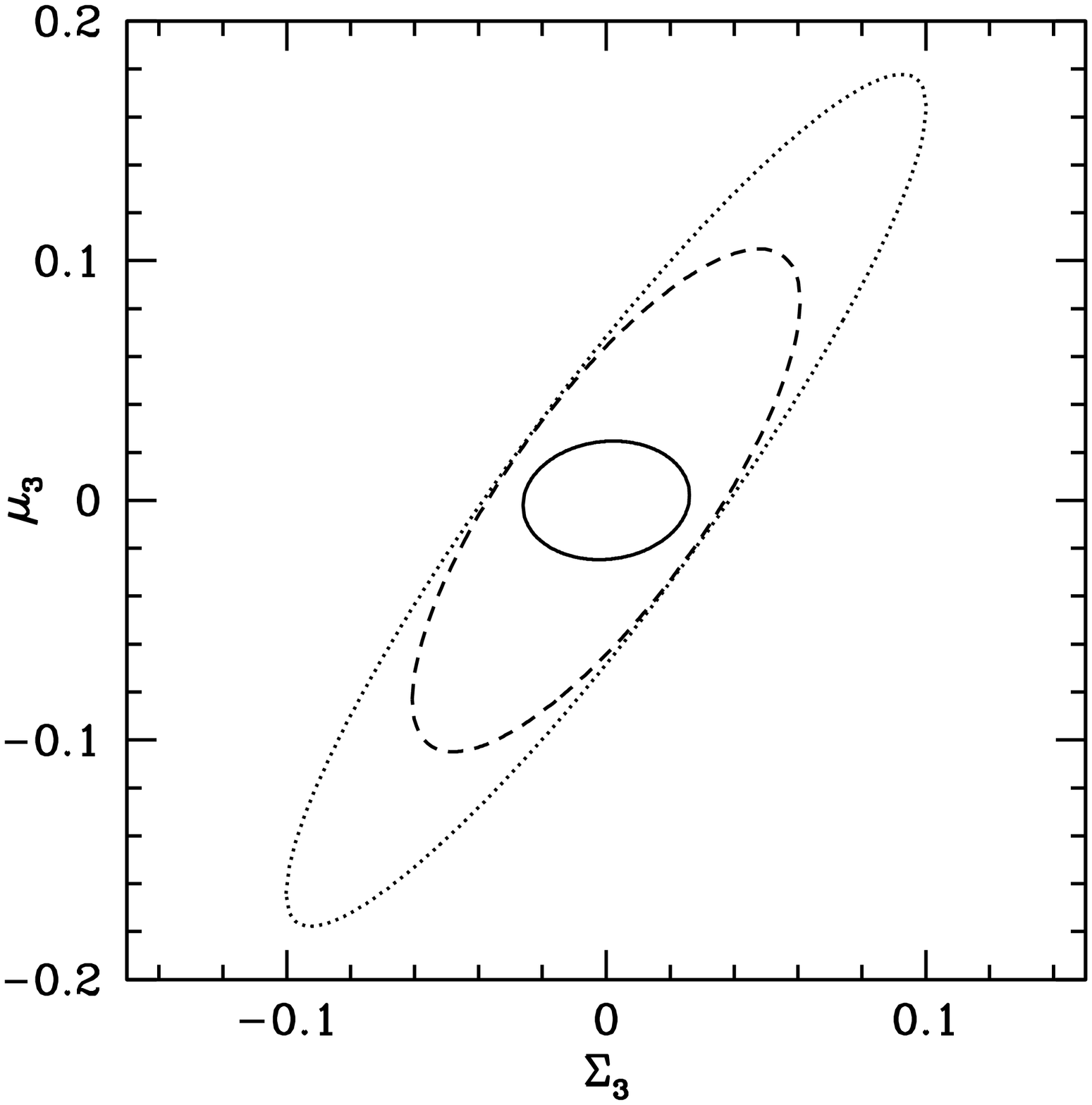}
\caption{Contour plots of $\Sigma_s$ and $\mu_s$ calculated from our Fisher forecast analysis.  The left panel shows  the $s=1$ case, while the right panel shows the $s=3$ case. Solid curves represent the scenario in which all other cosmological parameters except $\Sigma_s$ and $\mu_s$ are fixed, while the dashed curves represent the scenario in which $w=-1$ is assumed, and the dotted curves represent the scenario where all the cosmological parameters are allowed to vary including $w$.}\label{fig:future}
\end{figure*}

The left panel of Fig.~\ref{fig:current} shows our constraints on $\Sigma_1$ and $\mu_1$ for the $s=1$ model. In this particular model, any departures from the GR value of $\mu=0$ would appear at relatively high redshift and hence the effect on the density contrast $\delta$ is more prominent than for the $s=3$ model. One can see a strong degeneracy between  $\mu_1$ and $\Sigma_1$ in the WL constraints: if $\mu_1$ is large, the growth rate is enhanced, which can be compensated for by decreasing $\Sigma_1$. However, this degeneracy does not continue if $\Sigma_1$ drops below $-1$; since the WL observables depend only on the square of $\Sigma$, a large negative value of $\Sigma_1$ will actually enhance the lensing signal and can no longer compensate for enhanced growth due to large $\mu_1$. For negative $\mu_1$, the growth rate is suppressed and this can be compensated for by increasing $\Sigma_1$. In this case there is no upper bound for $\Sigma_1$, and no lower bound for $\mu_1$.

The right panel of Fig.~\ref{fig:current} shows the constraints on the parameters for the $s=3$ model. In this model, $\mu$ is now modified at relatively low redshifts, so the effect of varying $\mu_3$ on the lensing measurements is weaker than in the $s=1$ model above. As a consequence, the WL constraints now become more vertical in the $\Sigma_3- \mu_3$ plane as essentially only $\Sigma_3$ determines the WL signal.

Overall, Fig.~\ref{fig:current} shows the usefulness of using PV to break the $\Sigma-\mu$ degeneracy inherent in using just the WL measurements. For both models considered, the combined (blue) PV+WL contours are substantially tighter than those obtained from using either the PV or WL measurements alone.

In Table~\ref{table1}, we present our measurements of $\Sigma_s$ and $\mu_s$   for the combined WL and PV analysis, as well as measurements using the datasets separately. Again, we see significant improvement in the constraints on $\Sigma_s$ and $\mu_s$ when the two datasets are used together as illustrated in Fig. 1. For example, using both PV and WL data together, we obtain $\mu_1=-0.002 \pm 0.59$, $\Sigma_1 =0.11^{+0.59}_{-0.55}$ (for $s=1$) and $\mu_3=-0.08 ^{+2.27}_{-2.28}$, $\Sigma_3=0.17^{+0.88}_{-0.82}$ (for $s=3$) at the $95\%$ confidence level, which is fully consistent with the expectation for GR (i.e., $\mu_1=\mu_3=\Sigma_1=\Sigma_3=0$) although the errors are large, especially for the cubic model.

\section{Future forecasts}
\label{forecast}

In this section, we provide joint error forecasts on $\Sigma_s$ and $\mu_s$ from a future space--based DE mission such as Euclid \cite{Euclid}.

\subsection{Fisher matrix analysis}

The shear two-point functions and PV measurements depend not only on the MG parameters ($\Sigma_s,\mu_s$)  but also on the background cosmological parameters and the primordial power spectrum. We do not assume that these quantities are known, but instead assume that CMB data and supernova Type Ia  (SNIa) measurements will be available to help constrain them.  For the CMB power spectra, we include in our analysis the (unlensed) $C_l$ spectra of temperature--temperature, temperature--polarization and polarization--polarization and use the expected errors for the Planck survey \cite{Planck}. We include constraints from a deep space-based SNIa experiment \cite{Aldering} with observations of $3000$ supernovae with median $z\simeq0.8$, and assume that the SNIa evolution uncertainty is well understood. In the forecast, we treat ($\omega_b$, $\omega_c$, $\Theta_s$, $z_{reion}$, $n_s$, $A_s$,$w_0$, $w_a$, $\mu_s$, $\Sigma_s$) as free parameters with mean values of the WMAP7 best fit where the dark energy equation of state is modeled as $w(a)=w_0 + w_a (1-a)$ and $z_{reion}$ is the re-ionization epoch.

To calculate the expected errors on these parameters, we make a first--order Taylor expansion for the parameter dependences of all the observables (CMB, SN, cosmic shear two--point functions and PV). In this ``linear response" approximation, given the expected experimental errors on the power spectra, we can easily calculate the expected error covariance matrix as the inverse of the Fisher matrix. This ``linear response" approximation (using the first order Taylor expansion) can be improved with a careful choice of the parameters including $\Theta_S$~\cite{Kaplinghat:2003bh}. 

We consider a future WL survey such as Euclid \cite{Euclid} covering 20,000 square degrees with sufficient optical and infra--red sensitivity to yield a galaxy density of 35 galaxies per square arcminute \cite{Amara}. We also assume five redshift bins, with a spacing of $\Delta z=0.4$, and an underlying redshift distribution of $dn/dz \propto z^2 \exp\left[-(z/0.64\right)^{1.5}]$ \cite{Amara}. The contribution to a Fisher matrix from the shear--shear correlations is then given by
\be
F^\wl_{pp'}=\sum_{l,a,b,c,d}\frac{2l+1}{2}C_{l,p}^{\kappa\kappa\,ab}{\cal W}_l^{bc}C_{l,p'}^{\kappa\kappa\,cd}{\cal W}_l^{da},
\ee
where $C_l^{a,b}$ is the angular cross-power spectrum between redshift bin $a$ and $b$, $\cal W$ is the inverse of the total convariance matrix, and the subscript $p$ denotes differentiation with respect to a particular cosmological parameter~\cite{Song:2004tg}.

\begin{center}
\begin{table*}
\begin{tabular}{|c|l|cc|cc|}
\hline
\multicolumn{2}{|c|}{} & \multicolumn{2}{c|}{$s=1$} &\multicolumn{2}{c|}{$s=3$}\\
\multicolumn{2}{|c|}{} &  $\sigma(\Sigma_1)$  & $\sigma(\mu_1)$ & $\sigma(\Sigma_3)$  & $\sigma(\mu_3)$ \\
\hline
\multicolumn{2}{|c|}{Current constraint}   &  0.59  &  0.59 & 0.88 &  2.3  \\
\hline
 & Fix cosmological params.  & 0.0081 & 0.0046 & 0.017 & 0.014 \\
Future forecast & Assume only $w=-1$ & 0.035& 0.041& 0.040& 0.069\\
 & Allow $w$ to vary & 0.037 & 0.046 & 0.066 & 0.12 \\
\hline
\end{tabular}
\caption{The predicted errors on $\Sigma_s$ and $\mu_s$ for a future Euclid-like satellite mission compared to the current constraints obtained in this paper.}
\label{tab:error}
\end{table*}
\end{center}

Using spectroscopic redshift measurements, we are able to isolate the PV power spectra from the redshift-space power spectrum $P_g^{\rm obs}({\bf k})$ of a galaxy redshift survey. The latter is commonly modeled as~\cite{jackson}
\ba
 P_g^{\rm obs}({\bf k}) &=& \left[ P_{gg}^{}({\bf k})
                    + 2c^2P_{g\Theta}^{}({\bf k})
                    + c^4P_{\Theta\Theta}^{}({\bf k})\right]\nn\\
 &\times&F\left(k^2c^2\sigma_v^2(z)\right) ,
 \label{eq:pk_s1}
\ea
where $\Theta=\theta/aH$, $c$ is the cosine of the angle to the line of sight, $P_{gg}$ is the true galaxy power spectrum, $P_{\Theta\Theta}$ is the peculiar velocity power spectrum and $P_{g\Theta}$ is the galaxy-velocity cross spectrum. The separation of $P_{gg}({\bf k})$ and $P_{\Theta\Theta}({\bf k})$ is made possible using the angular dependence of $P_g^{\rm obs}({\bf k})$.

In order to obtain cosmological constraints from the peculiar velocities, we need to understand the error covariance matrix of the observables. We estimate this using the Fisher matrix \cite{White:2008jy} 
\ba
\lefteqn{F^{\rm obs}_{\alpha\beta}(k_i,z_j) = \int_{k_i^{\rm min}}^{k_i^{\rm
      max}}\frac{k^2dk}{2(2\pi)^2}\int^1_{-1} dc\nn\ V_{\rm eff}(k,c,z_j)}&&\\
&\times&\frac{\partial \ln P_{g}^{\rm obs}(k,c,z_j)}{\partial
  p_{\alpha}} \frac{\partial \ln P_{g}^{\rm obs}(k,c,z_j)}{\partial
  p_{\beta}} ,
\label{eq:fish_gg_tt}
\ea
where $\alpha$ and $\beta$ both run from 1 to 2 and denote $P_{gg}$ and $P_{\Theta\Theta}$ respectively. Note that in the linear regime $P_{g\Theta}$ is simply the square root of the product of $P_{gg}$ and $P_{\Theta\Theta}$, so we take this to be the case in this work. The effective volume $V_{\rm eff}^j$ in each
redshift bin $j$ is given by
\be
V_{\rm eff}(k_i,c,z_j) =
\left[\frac{n^jP_{g}^{\rm obs}(k_i,c,z_j)}{n^jP_{g}^{\rm obs}(k_i,c,z_j)+1}\right]^2
V_{\rm survey}(z^j) ,
\ee
where $n^j$ is the shot noise term coming from the finite galaxy density, and $V_{\rm survey}(z_j)$ is the survey volume in a given redshift bin. On large scales, the cosmic variance term dominates over the shot noise term and $V_{\rm  eff}(k_i,\mu,z_j)$ is nearly identical to $V_{\rm survey}(z_j)$. For our estimation, we consider a 20000 sq deg Euclid-like survey with H-$\alpha$ emitter number density as a function of redshift given by \cite{geach} (Table 2, limiting flux $4\times10^{-16}$ erg s$^{-1}$ cm$^{-2}$).

Following the above estimation of $P_{gg}$, $P_{\Theta\Theta}$ and their errors, we can then constrain the cosmological parameters from the decomposed $P_{\Theta\Theta}$. (Here we elect not to use $P_{gg}$ on account of the bias). The Fisher matrix for cosmological parameters, using the peculiar velocity spectrum, can be written as \cite{Song:2010ia} 
\be\label{eq:fish_gt}
F^{\vp}_{pp'}=\sum_{ij}
\frac{\partial P_{\Theta\Theta}}{\partial p}
\frac{1}{(F^{-1\,\rm obs})_{22}(k_i,z_j)}
\frac{\partial P_{\Theta\Theta}}{\partial p'},
\ee
where {\it i} denotes $k$ bins up to $k=0.1\hompc$, and {\it j}
denotes redshift bins from $z=0$ to 2 with spacing $\Delta z=0.2$.

\subsection{Future forecasts}
In this subsection, we study the constraints on the MG parameters in three different scenarios: 1) we only vary the parameters $\Sigma_s$, $\mu_s$, fixing all other cosmological parameters; 2) we vary all cosmological parameters with the assumption of a flat $\Lambda$CDM expansion history; 3) we vary all cosmological parameters including $w$ using $w(a)=w_0+w_a(1-a)$. In each case, we again consider the $s=1$ and $s=3$ models.

Fig.~{\ref{fig:future}} shows constraints on $\Sigma_s$ and $\mu_s$ in the three scenarios above. Again, thanks to the complementarity of WL and PV measurements, the constraints on these MG parameters are very tight (solid contours) if we fix the other cosmological parameters. However, if we marginalise over the uncertainties in the other cosmological parameters, then the constraints on $\Sigma_s$ and $\mu_s$ are degraded by an order of magnitude: the dotted contours in Fig.~{\ref{fig:future}} represent the full marginalization over other cosmological parameters allowing $w$ to vary, while the inner dashed contours represent the $w=-1$ case.

In Table 2, we present the expected constraints on $\mu_s$ and $\Sigma_s$ for the three scenarios discussed above. For comparison, we also provide in the table the current constraints on these MG parameters and even in the most conservative case (allowing $w$ to vary), the future satellite constraints should be at least 20 times better. It is interesting to note the difference between the $s=1$ and the $s=3$ model. As $s$ decreases, the expected departure from GR starts at an earlier epoch thus resulting in a stronger constraint.

Comparing the current constraints in Fig.~\ref{fig:current} and the future forecasts in Fig.~\ref{fig:future}, we notice a change in the orientation of the $\Sigma_s-\mu_s$ contours.
Namely, in the current constraints, there is an anti--correlation between $\Sigma_s$ and $\mu_s$, while for the future forecasts, this turns into a positive correlation. We note that without inclusion of the PV information there is still an anti--correlation between $\Sigma_s$ and $\mu_s$ in the forecast as well, so it is the inclusion of PV which is changing the orientation.

This change in orientation is due to the effect of marginalising over other cosmological parameters. To better understand this effect, let us examine the effect of marginalizing over $\omega_m$.  When we examine the covariance between different cosmological parameters for a PV experiment, we find that both $\Sigma_s$ and $\mu_s$ are strongly anti-correlated with $\omega_m$, and to a lesser extent with $w_0$, $n_s$, $\theta_S$, $\omega_b$ and $A_S$; this means, for instance, that an increase in $\mu$ can be compensated for by a decrease in $\omega_m$. Equally, if one decreases $\omega_m$, the WL data can be fitted if one also increases $\Sigma$. As a result, given precise measurements for both PV and WL, $\Sigma_s$ and $\mu_s$ can become positively correlated, as they may both change in the same sense to oppose a change in $\omega_m$ in the opposite sense. However, if the PV data are sufficiently noisy, the anti-correlation of $\mu$ and $\Sigma$ exhibited in the WL constraints alone (see Fig.~\ref{fig:current}) will dominate.

%%%%%%%%%%%%%%%%%%%%%%%%%%%%%%%%%
\section{Towards model independent constraints on MG parameters}
\label{independent}

As we saw in section~\ref{current} and section~\ref{forecast}, constraints on $\mu_s$ (and to a lesser degree $\Sigma_s$) depend strongly on the assumed value of $s$. One may wonder if it is possible to extract any model-independent information about $\mu$ from the current PV and WL data. In principle, one can bin $\mu$ into a large number of narrow redshift bins and find their best constrained uncorrelated linear combinations -- the so-called eigenmodes \cite{Huterer:2002hy,Crittenden:2005wj,Zhao:2009fn}. In practice, this would be a numerically challenging project, given the large range of $z$ over which variations of $\mu$ can affect the growth. However, one would expect that with PV and WL measurements at a single redshift $z^*$, one would constrain just one of the eigenmodes, corresponding to a net change in $\mu$ over $0<z<z^*$. The tightness of the constraint would depend on how early the departures from $\mu=1$ are allowed to start, because an earlier change in $\mu$ results in a larger change in the growth factor throughout $z$.

\begin{figure}[t]
\includegraphics[width=9cm]{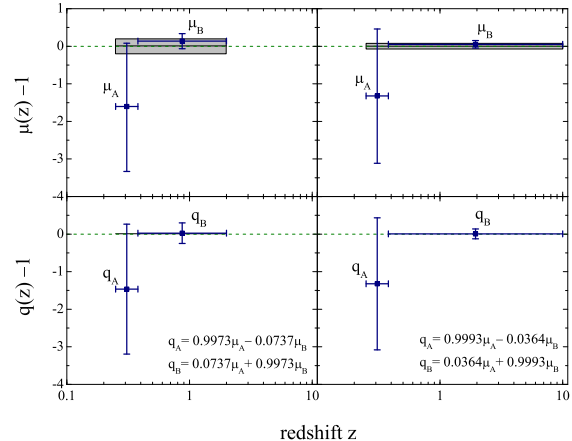}
\caption{Constraints from PV data on $\mu$ as a function of redshift. We show the results of using 2 redshift bins, with $\mu=\mu_A$ within [$z_1,z_2$] and $\mu=\mu_B$ within [$z_2,z_s$]. It is assumed that $mu=1$ above $z_s=2$ (left) and above $z_s=10$ (right).
In addition we display (grey box) results for constant $\mu$ below redshift $z_s=2$ (left) and below $z_s=10$ (right) when PV is combined with CMB shift parameter and SNe. The lower panels show the linear combination of $\mu_A$ and $\mu_B$ that diagonalise the covariance matrix.}
\label{fig:2bin}
\end{figure}

To demonstrate this point, we try fitting $\mu$ (which can be any constant over a wide bin) between $z_s$ and $z=0$, and then setting $\mu=1$ for $z>z_s$. Note that this is not quite the same as the $s=0$ case in Eqn.~(\ref{parametrization}), as in this equation we do not explicitly require $\mu$ to be exactly unity above a certain redshift. The results for the choices of $z_s=2$ and $z_s=10$ are shown as shaded boxes in the top right and left panels respectively of Fig.~\ref{fig:2bin}. As expected, the constraint on the constant bin is much stronger in the $z_s=10$ case.

However, the analysis of the SDSS DR7 LRGs provides measurements of the peculiar velocities of galaxies at more than one redshift, i.e. we have measurements at $z_1=0.25$ and $z_2=0.38$. Therefore, one could ask if it is possible to obtain additional information about the variation of $\mu$ between $z_2$ and $z_1$ independent of the assumed value of $z_s$. To address this, we examine $\mu$ in two bins, namely $\mu_A$ in [$z_1,z_2$] and $\mu_B$ in [$z_2,z_s$], with $\mu=1$ for $z>z_s$.

The constraints on these two bins are shown again in Fig.~\ref{fig:2bin} for the two cases of $z_s=2$ and $z_s=10$.  As expected, the constraint on $\mu$ is tight in the higher redshift bin, with the error bar being much smaller for the $z_s=10$ case. On the other hand,  the errors on the low redshift bin are much larger. Interestingly, however, they are effectively independent of the choice of $z_s$, as is the correlation coefficient between the two bins.

By diagonalising the covariance matrix we find the uncorrelated linear combinations of the two bins to be $q_A=0.9973 \mu_A -0.0737 \mu_B$ and $q_B=0.0737 \mu_A + 0.9973 \mu_B$ for the $z_s=2$ case and  $q_A=0.9993 \mu_A -0.0364 \mu_B$ and $q_B=0.0364 \mu_A + 0.9993 \mu_B$ for the $z_s=10$ case. One can see from Fig.~\ref{fig:2bin} that the error on the better constrained eigenmode depends strongly on $z_s$, while the error on the second eigenmode stays the same. This demonstrates that having measurements at multiple redshifts can produce constraints on the variation of $\mu$ at low redshifts that are independent of its assumed time-dependence at high redshifts. Finally, we note that the values for $\mu$ found in the two bins do not show any deviation from the GR predictions.

\section{Conclusion}
\label{conclusion}

In this paper, we have explored the complementarity of weak lensing and peculiar velocity measurements to obtain better constraints on the modified gravity parameters $\mu$ and $\Sigma$, as defined in Eqns.~(\ref{eq:MG_para1}) and (\ref{eq:MG_para2}). Peculiar velocity measurements are only sensitive to variations of $\mu$, thus breaking the degeneracy between $\mu$ and $\Sigma$ found when using just the weak lensing measurements. Using a simplified parametrization for $\mu$ and $\Sigma$ (see Eqn.~\ref{parametrization}), we have derived the first constraints on $\mu$ and $\Sigma$ using weak lensing measurements and peculiar velocity measurements (from CFHTLS and SDSS respectively). Assuming a flat $\Lambda$CDM background cosmology, so that here any modifications to gravity are probed by structure growth, the best constraints we find using both datasets together are $\mu_1=-0.002 \pm 0.59$ and $\Sigma_1 =0.11^{+0.59}_{-0.55}$ for $s=1$, and $\mu_3=-0.08 ^{+2.27}_{-2.28}$ and $\Sigma_3=0.17^{+0.88}_{-0.82}$ for $s=3$ ($95\%$ confidence limits). These results are consistent with the expectation from General Relativity, namely $\mu_1=\mu_3=\Sigma_1=\Sigma_3=0$. As shown in Table 1, these constraints are much worse if the WL and PV data are used separately, in which case some of the parameters become unconstrained.

We also performed a Fisher error forecast for a space--based Dark Energy mission like Euclid. For example, assuming a flat $\Lambda$CDM cosmological background, we predict that the constraints on the modified gravity parameters are improved by at least an order of magnitude or more, e.g. $\sigma(\mu_1)=0.037$, $\sigma(\Sigma_1)=0.032$ and $\sigma(\mu_3)=0.068$, $\sigma(\Sigma_3)=0.038$ (see Fig.~\ref{fig:future} and Table~\ref{tab:error} for the full details). It is interesting to note that the expected precision on these errors is comparable in size to the present--day  uncertainties on cosmological parameters like $w$ (approximately known to 10\% today) and $\Omega_m$ ($\sim5\%$ error from present observations). Therefore, a Euclid--like mission will deliver outstanding constraints on modified gravity as well as improving errors on the background cosmological parameters.

That said, we found that the constraints on $\mu$ depend strongly on the assumed time variation of this parameter, i.e. $s$ in Eqn.~(\ref{parametrization}). We have therefore proposed a method for extracting
model--independent information about $\mu$ by examining $\mu$ in different redshift bins. For the SDSS DR7 PV measurements, we have used two bins and are able to obtain a model--independent constraint on $\mu$ at low redshifts which is again consistent with expectations for GR (see Fig.~\ref{fig:2bin}). This demonstrates that having PV measurements at multiple redshift intervals can produce a constraint on the variation of $\mu$, at low redshifts, which is independent of the assumed time dependence at high redshifts. This technique will become increasingly important in the next few years as new peculiar velocity measurements are published from surveys like WiggleZ \cite{Wigglez}, BOSS and VIPERS.

%***Check journal entries - ref as well as astroph, consistency

\acknowledgments
GZ, DB and KK are supported by STFC grant ST/H002774/1, and DB and RCN are supported by STFC grant ST/F002335/1. DB and KK are also supported by RCUK. KK acknowledges support from the European Research Council and the Leverhulme Trust. LP is supported by an NSERC Discovery Grant. YSS thanks Korea Institute for Advanced Study for providing computing resources (KIAS linux cluster system) for this work. LP thanks Portsmouth ICG for their hospitality during the course of this work.

\end{document}